\documentclass[12pt]{article}
\usepackage{graphicx}

\newcommand{\be}{\begin{equation}}
\newcommand{\ee}{\end{equation}}

\begin{document}

\noindent {\small CITUSC/01-008 \hfill \hfill hep-th/0103209 \newline
IC/2001/19 \hfill \newline
FIAN/TD/08-01 \hfill }


{\vskip-0.8cm}

\begin{center}
{\Large \textbf{Noncommutative o}}$_{\star }${\Large \textbf{(N) and usp}}$%
_{\star }${\Large \textbf{(2N) Algebras \\[0pt]
And The Corresponding Gauge Field Theories}}

{\vskip0.4cm}

\textbf{I. Bars}$^{a}$\textbf{, M.M. Sheikh-Jabbari}$^{b}$ \textrm{and}
\textbf{M.A.Vasiliev}$^{c}$

{\vskip0.3cm}

$^{a)}$\textsl{CIT-USC Center for Theoretical Physics \& Department of
Physics}

\textsl{University of Southern California,\ Los Angeles, CA 90089-2535 USA}

\texttt{bars@usc.edu}

{\vskip0.3cm}

$^{b)}$\textsl{The Abdus Salam International Center for Theoretical Physics}

\textsl{Strada Costiera 11, 34014 Trieste, Italy}

\texttt{jabbari@ictp.trieste.it}

{\vskip0.3cm}

$^{c)}$\textsl{I.E. Tamm Department of Theoretical Physics, Lebedev Physical
Institute}

\textsl{Leninsky Prospect 53, 117954, Moscow, Russia}

\texttt{vasiliev@lpi.ru}

{\vskip0.5cm} \textbf{Abstract}
\end{center}

The extension of the noncommutative u$_{\star }(N)$ Lie algebra to
noncommutative orthogonal and symplectic Lie algebras is studied. Using an
anti-automorphism of the star-matrix algebra, we show that the u$_{\star
}(N) $ can consistently be restricted to o$_{\star }(N)$ and usp$_{\star
}(N) $ algebras that have new mathematical structures. We give explicit
fundamental matrix representations of these algebras, through which the
formulation for the corresponding noncommutative gauge field theories are
obtained. In addition, we present a D-brane configuration with an
orientifold which realizes geometrically our algebraic construction, thus
embedding the new noncommutative gauge theories in superstring theory in the
presence of a constant background magnetic field. Some algebraic
generalizations that may have applications in other areas of physics are
also discussed.\newpage

\baselineskip=18pt

\section{Introduction}

Noncommutative (NC) spaces have been shown to arise from string theory \cite
{{CDS}}. More precisely the world-volume coordinates of D$p$-branes living
in a constant B-field background turn out to be noncommuting
\begin{equation}
\left[ x^{\mu },x^{\nu }\right] _{\star }=i\theta ^{\mu \nu },\quad \mu ,\nu
=0,1,\cdots ,2n;\quad 2n\leq p+1\ ,  \label{ncspace}
\end{equation}
where $\theta ^{\mu \nu }$ is a $2n\times 2n$ real antisymmetric matrix $%
\theta ^{\mu \nu }=-\theta ^{\nu \mu }$, which is a certain function of the
background B-field and metric \cite{SW}. As a result, the low energy
effective theory of the open strings attached to such NC branes becomes a NC
gauge theory \cite{{SW},{HD}}. The case of NC u(N) gauge theory is well
understood, but despite previous attempts \cite{{Bonora},{Wess}}, the cases
of o$(N)$ and usp$\left( 2N\right) $ have escaped a full understanding. In
particular, problems of non-renormalizability have arisen \cite{bonor} with
the previous definition of these theories. In this paper we will introduce a
new definition of the noncommutative algebras o$_{\star }\left( N\right) $
and usp$_{\star }\left( 2N\right) $ and construct the corresponding gauge
field theories. We will also construct the geometry of the D-branes that
give rise to these gauge theories. The resulting o$_{\star }\left( N\right) $
and usp$_{\star }\left( 2N\right) $ \ gauge theories look rather different
than the previous suggestions.

The main ingredient in the construction at the algebraic level is an
anti-automorphism of the noncommutative space. Geometrically, this is
related to an orientifold of a new type which had not been considered in
brane constructions of gauge theories so far. The new anti-automorphism
overcomes certain conceptual problems that were encountered in the previous
attempts to construct o$_{\star }\left( N\right) $ and usp$_{\star }\left(
2N\right) $ \ gauge theories.

The star product between two functions $f\left( x\right) $ and $g\left(
x\right) $ over the NC space is defined by the Moyal star product
\begin{equation}  \label{star}
f(x)\star g(x)=\left. \exp \left( \frac{i}{2}\theta ^{\mu \nu }\frac{%
\partial ^{2}}{\partial x_{1}^{\mu }\partial x_{2}^{\nu }}\right)
f(x_{1})g(x_{2})\right| _{x_{1}=x_{2}=x}.
\end{equation}
This associative product defines an algebra $\mathcal{A}$ in the space of
functions on NC space. It is clear that the functions that belong to $%
\mathcal{A}$ are generally complex functions since the star product
necessarily introduces the complex number $i$. Hence, in all of our
discussion it will be understood that all non-commutative functions are
generically complex; for example, locally, one may think of them as a power
series in real coordinates $x^{\mu }$ with complex coefficients. We will
define complex conjugation $\bar{f}\left( x\right) $ to mean the complex
conjugation of the coefficients in the power series.

The star commutator that occurs in Eq.(\ref{ncspace}) is defined by $\left[
f,g\right] _{\star }=f(x)\star g(x)-g(x)\star f(x).$ One may compute some
examples of products which will be useful in the discussion below
\begin{eqnarray}
x^{\mu }\star x^{\nu } &=&x^{\mu }x^{\nu }+\frac{i}{2}\theta ^{\mu \nu },
\label{two} \\
\left[ x^{\mu },x^{\nu }\right] _{\star } &\equiv &x^{\mu }\star x^{\nu
}-x^{\nu }\star x^{\mu }=i\theta ^{\mu \nu },  \label{twop} \\
x^{\mu }\star f\left( x\right) &=&\left( x^{\mu }+\frac{i}{2}\theta ^{\mu
\nu }\partial _{\nu }\right) f\left( x\right) , \\
f\left( x\right) \star x^{\mu } &=&\left( x^{\mu }-\frac{i}{2}\theta ^{\mu
\nu }\partial _{\nu }\right) f\left( x\right) , \\
x^{\mu }\star x^{\nu }\star x^{\lambda } &=&x^{\mu }x^{\nu }x^{\lambda }+%
\frac{i}{2}\theta ^{\mu \nu }x^{\lambda }+\frac{i}{2}\theta ^{\mu \lambda
}x^{\nu }+\frac{i}{2}\theta ^{\nu \lambda }x^{\mu }.  \label{three}
\end{eqnarray}

The approach in this paper is rather general and could be used in some other
models as well. In particular, it was applied to the classification of
various types of higher spin algebras compatible with the dynamics of higher
spin gauge fields in $AdS_{4}$ \cite{4} and $AdS_{3}$ \cite{3}. The higher
spin algebras contain subalgebras that correspond to ordinary Yang-Mills
symmetries (i.e. spin one) of unitary, orthogonal and symplectic types.
Since the higher spin gauge theories are formulated in terms of auxiliary
noncommutative spaces with spinor coordinates, the formalism is in many
respects analogous to that of the non-commutative Yang-Mills theory (see
\cite{hsr} for reviews and more references on the higher spin gauge theory).
Also, examples of algebras that have some relation to those introduced in
this paper for the case of noncommutative plane were discussed in other
contexts and formalisms, in particular see \cite{FV} in the context of the
higher spin theories for the case of hyperbolic geometry, and \cite{FFZ} for
the toric case.

The paper is organized as follows. In section 2 we present the problem,
describe previous attempts, point out some difficulties, and then present an
explicit construction of the algebras o$_{\star }\left( N\right) $ and usp$%
_{\star }\left( 2N\right) $ based on an anti-automorphism of the algebra $%
\mathcal{A}$. The gauge theories follow naturally once the new algebras are
defined. In section 3 we present the D-brane geometry that leads to these
gauge theories. In section 4 we present a more formal exposition that
generalizes the mathematical setup to a broader range of structures that
have applications to other problems in physics. In section 5 we give a
summary and discuss open problems.

\section{Algebraic structure of o$_{\star }(N)$ and usp$_{\star }(2N)$}

The classical su$\left( N\right) ,$ so$\left( N\right) ,$ and usp$\left(
2N\right) $ Lie algebras can be defined through their fundamental matrix
representations. Namely, one notes that under ordinary matrix commutators
the following sets of matrices form Lie algebras: su$\left( N\right) :$ $%
N\times N$ antihermitian traceless matrices $h=-h^{\dagger }$ over complex
numbers, so$\left( N\right) :$ $N\times N$ antisymmetric matrices $a^{t}=-a$
over real numbers (where $t$ denotes matrix transposition), and usp$\left(
2N\right) :$ $2N\times 2N$ antihermitian matrices that satisfy $%
s^{t}=-CsC^{-1}$(equivalently $\left( Cs\right) ^{t}=Cs,$ is symmetric)
where
\begin{equation}
C=\left(
\begin{array}{cc}
0 & 1_{N} \\
-1_{N} & 0
\end{array}
\right) .  \label{C}
\end{equation}
Of course, analytic continuation to various real forms are also possible,
but since this is a trivial point in our discussion we will not dwell on it
presently. Ordinary gauge theories based on these Lie algebras contain local
gauge fields $A_{\mu }\left( x\right) $ that are matrices $\left( A_{\mu
}\left( x\right) \right) _{\,\,j}^{i}$ of the forms $h,a,s$ for su$\left(
N\right) $, so$\left( N\right) ,$ usp$\left( 2N\right) $ respectively. The
coordinates $x^{\mu }$ belong to a spacetime manifold $M.$

In noncommutative gauge theories the spacetime coordinates are replaced by
noncommuting coordinates as in Eq.(\ref{ncspace}). The gauge fields $A_{\mu
}\left( x\right) $ belong to \textit{Mat}${}_{N}\otimes \mathcal{A}\quad$
that is matrices whose matrix elements $\left( A_{\mu }\left( x\right)
\right) _{\,\,j}^{i}$ are noncommutative functions that belong to $\mathcal{A%
}$.

When the entries of the matrices are elements of $\mathcal{A}$, the product
involves not only matrix product but also the star product. This deformation
in the product destroys the simple matrix closure of the classical Lie
algebras, and a new definition must be introduced to find the sets of
matrices \textit{Mat}${}_{N}\otimes \mathcal{A}\quad$ that close to form the
noncommutative versions of the classical Lie algebras. To see the point
clearly, suppose we start naively by commuting two so$\left( 2\right) $
matrices filled with noncommutative functions $f\left( x\right) $ and $%
g\left( x\right) $
\begin{equation}
\left(
\begin{array}{cc}
0 & f \\
-f & 0
\end{array}
\right) \star \left(
\begin{array}{cc}
0 & g \\
-g & 0
\end{array}
\right) -\left(
\begin{array}{cc}
0 & g \\
-g & 0
\end{array}
\right) \star \left(
\begin{array}{cc}
0 & f \\
-f & 0
\end{array}
\right) =\left(
\begin{array}{cc}
\left[ g,f\right] _{\star } & 0 \\
0 & \left[ g,f\right] _{\star }
\end{array}
\right)
\end{equation}
If $f,g$ were commutative functions the result would have been zero,
corresponding to the closure of ordinary so$\left( 2\right).$ However, with
noncommutative functions $f,g$ the algebra does not close since the result
is a different form of matrix. Therefore, we need to find the new sets of
matrices over the noncommutative algebra $\mathcal{A}$ that close. We will
name the new noncommutative Lie algebras u$_{\star }\left( N;n\right) ,$ o$%
_{\star }\left( N;n\right) $ and usp$_{\star }\left( 2N;n\right),$ where $n$
is the number of pairs of conjugated coordinates in (\ref{ncspace}). In the
rest of the paper we will skip the label $n$, however.

The case of u$_{\star }\left( N\right) $ is well understood and extensively
used in the literature. Nevertheless we will include it in our discussion in
order to provide a background toward the noncommutative o$_{\star }\left(
N\right) $ and usp$_{\star }\left( 2N\right) .$

To see what is needed, it is useful to think of the deformation introduced
by the star product as being similar to replacing the algebra $\mathcal{A}$
by the algebra of quantum operators or the algebra of matrices. Then one
encounters the same closure problem. To find the correct sets of matrices
\textit{Mat}${}_{N}\otimes \mathcal{A}\quad$ that close, the definition of
hermitian conjugation or transposition, which entered in the definition of su%
$(N)$, so$(N)$ or usp$(2N)$, would have to be extended to the operator or
matrix entries as well. Without such a definition the set into which the Lie
algebra would close is not specified. The same idea must be applied to the
noncommutative algebra $\mathcal{A}$.

Let us consider a map $\rho $ (defined explicitly later) such that, when
acting on the elements of $\mathcal{A}$\ it has the property
\begin{equation}
\rho \left( \left( f\star g\right) \left( x\right) \right) =\left( \rho
\left( g\right) \star \rho \left( f\right) \right) \left( x\right) ,\ \ \ \
\rho (\rho (f(x)))=f(x)\ ,  \label{rhomap}
\end{equation}
shared by hermitian conjugation and transposition. The reversal of the
orders in the star product is the crucial property. Then, we combine
ordinary matrix hermitian conjugation or transposition with the map $\rho $
to define an antiautomorphism for the algebra \textit{Mat}${}_{N}\otimes
\mathcal{A}\quad $. Under the combined operation we demand antihermitian
matrices, antisymmetric matrices etc. to define the Lie algebras we are
seeking. The main issue is to find an explicit form of the map $\rho $ that
works as desired in our context.

\subsection{u$_{\star }\left( N\right) $}

Hermitian conjugation on $\mathcal{A}$ is taken as the standard complex
conjugation of a complex function
\begin{equation}
\left( f\left( x\right) \right) ^{\dagger }=\bar{f}\left( x\right) \,,
\label{hc}
\end{equation}
defined by changing all $i$ into $-i.$ This definition is consistent with
the star product provided the order of the factors in the star product is
reversed
\begin{equation}
\left( f\left( x\right) \star g\left( x\right) \right) ^{\dagger }=\left(
g\left( x\right) \right) ^{\dagger }\star \left( f\left( x\right) \right)
^{\dagger }=\bar{g}\left( x\right) \star \bar{f}\left( x\right) .
\end{equation}
This is similar to the hermitian conjugation of the product of quantum
operators or matrices. The right hand side indeed gives the ordinary complex
conjugate of $f\left( x\right) \star g\left( x\right) ,$ including the sign
change of the $i$ introduced by the star product. This can be proven
generally from the definition of the star product in Eq.(\ref{star}):
changing the order of the functions on the right hand side of Eq.(\ref{star}%
) is equivalent to changing the sign of $\theta ^{\mu \nu },$ which is
equivalent to changing the sign of $i$ introduced by the star product. It is
perhaps useful to the reader to verify explicitly that the definition works
correctly on some explicit products such as those listed in Eqs.(\ref{two}-%
\ref{three}). For example, by using $x^{\mu }\star x^{\nu }=x^{\mu }x^{\nu }+%
\frac{i}{2}\theta ^{\mu \nu }$ we can evaluate the hermitian conjugation in
two ways, first, by complex conjugation of the right hand side, and second,
by applying the interchange rule to the star product, thus noting that the
result is the same ordinary function of $x,\theta $
\begin{eqnarray}
\left( x^{\mu }\star x^{\nu }\right) ^{\dagger } &=&x^{\mu }x^{\nu }-\frac{i%
}{2}\theta ^{\mu \nu }, \\
\left( x^{\mu }\star x^{\nu }\right) ^{\dagger } &=&x^{\nu }\star x^{\mu
}=x^{\mu }x^{\nu }-\frac{i}{2}\theta ^{\mu \nu },
\end{eqnarray}
where we have used $\theta ^{\nu \mu }=-\theta ^{\mu \nu }.$

To define u$_{\star }\left( N\right) $ we combine this definition of
hermitian conjugation of noncommutative functions with ordinary matrix
hermitian conjugation. For the combined hermitian conjugation we require
antihermitian matrices. Thus, u$_{\star }(N$) is defined by $N\times N$
matrices whose entries are noncommutative complex functions that satisfy $%
\left( \left( h\left( x\right) \right) ^{\dagger }\right) _{\,\,j}^{i}=-\bar{%
h}_{\,\,\,i}^{\,j}\left( x\right) .$ The matrices that satisfy these
relations have the form
\begin{equation}
u_{\star }(N):\quad h_{\,\,j}^{i}\left( x\right) =\left(
\begin{array}{cccc}
ih_{11}\left( x\right) & h_{12}\left( x\right) & \cdots & h_{1N}\left(
x\right) \\
-\overline{h_{12}}\left( x\right) & ih_{22}\left( x\right) & \cdots &
h_{2N}\left( x\right) \\
\vdots & \vdots & \ddots & \vdots \\
-\overline{h_{1N\left( x\right) }}\left( x\right) & -\overline{h_{2N}}\left(
x\right) & \cdots & ih_{NN}\left( x\right)
\end{array}
\right)  \label{H}
\end{equation}
where on the diagonal $ih_{ii}\left( x\right) $ are purely imaginary
functions, while the off-diagonal $h_{ij}\left( x\right) $ are complex
functions. (Note that the matrix trace cannot be subtracted since traceless
matrices do not close under the combined star-matrix commutation relations.
In particular u$_{\star }\left( 1\right) $ is non-Abelian, in contrast to
commutative u$\left( 1\right) $ which is Abelian.) Such matrices close to
form a Lie algebra. Thus, for $h_{1}$ and $h_{2}$ in the set, the
matrix-star commutator results in another $h_{3}$ in the set
\begin{equation}
\left( \left[ h_{1},h_{2}\right] _{\star }\right) _{\,\,\,j}^{i}\equiv
\left( h_{1}\star h_{2}-h_{2}\star h_{1}\right) _{\,\,\,j}^{i}=\left(
h_{3}\right) _{\,\,\,j}^{i}.  \label{closeUn}
\end{equation}

Gauge fields $A_{\mu }\left( x\right) $ based on u$_{\star }(N)$ are
matrices that have the same matrix form as Eq.(\ref{H}). The corresponding
gauge field strength $G_{\mu \nu }$ is also a similar matrix thanks to the
closure property
\begin{equation}  \label{curv}
\left( G_{\mu \nu }\right) _{\,\,\,j}^{i}=\left( \partial _{\mu }A_{\nu
}-\partial _{\nu }A_{\mu }+[A_{\mu },A_{\nu }]_{\star } \right)
_{\,\,\,j}^{i}\,.
\end{equation}
Gauge transformation parameters $\left( h\left( x\right) \right)
_{\,\,\,j}^{i}$ are also similar matrices. Under such gauge transformations $%
G_{\mu \nu }$ has the usual properties, again thanks to closure
\begin{equation}
\delta A_{\mu }=\partial _{\mu }h+\left[ A_{\mu },h\right] _{\star },\quad
\delta G_{\mu \nu }=\left[ G_{\mu \nu },h\right] _{\star }.
\end{equation}
So, in the case of u$_{\star }(N)$ the algebraic structure is very similar
to that of matrices.

To define an invariant action, the analog of the trace of a matrix is
introduced. This must include trace over the matrix as well as over the
noncommutative functions. It is well known that the integral over the $2n$
dimensional noncommutative space (for well behaved integrable functions) is
the correct definition. Thus, the invariant action for u$_{\star }(N)$ gauge
theory is
\begin{equation}
S=\int d^{p+1}x\,Tr\left( G_{\mu \nu }G^{\mu \nu }\right) ,  \label{act}
\end{equation}
where $Tr$ is the matrix trace.

\subsection{o$_{\star }\left( N\right) $}

In \cite{Bonora} a candidate for the map $\rho $ was suggested, and it was
called the $r$-map. It required considering the elements of the algebra \ $%
\mathcal{A}$ not simply as functions of spacetime $x^{\mu },$ but also as
functions of the non-commutativity parameter $\theta ,$ regarded as a
variable $f\left( x;\theta \right) $. Then the $r$-map was defined as $%
\left( f\left( x;\theta \right) \right) ^{r}=f\left( x;-\theta \right) .$
The authors showed that this definition leads to a Lie algebra which may be
called NC o$\left( N\right) ,$ and also presented a D-brane geometry that
seemed compatible with a gauge theory based on this Lie algebra. The brane
geometry required a $B$ field that was not a constant, but needed to be a
step-function in the directions transverse to the orientifold. The main
problem in this approach, as anticipated by the authors, is the fact that
the $\theta $ that appears in the low energy theory is a constant, not a
variable. When regarded as a variable, one may expand $Mat\left( \mathcal{A}%
\right) $ in a power series in $\theta $ and identify the matrix
coefficients of the series as independent generators of an infinite
dimensional algebra. This requires an infinite set of gauge fields, one for
each power of $\theta $; but in string theory there is only one set of gauge
fields. The authors speculated that this infinite set could be related to a
single set of gauge fields through the Seiberg-Witten map \cite{SW}.
However, one should anticipate purely on algebraic grounds that unless the
infinite set of gauge fields are independent, the corresponding gauge theory
is likely to be non-renormalizable. Indeed, signals that this is a problem
have already been reported in \cite{bonor}.

We will take here a different approach by defining a new explicit map $\rho $
for a fixed value of $\theta .$ In our case the map $\rho $ involves only
the spacetime coordinates $x^{\mu }.$ To define o$_{\star }\left( N\right) $
we introduce the analog of transposition for the non-commutative functions.
We will introduce two such transpositions denoted by $t_{1}$ and $t_{2}$
(candidates for $\rho $) as follows
\begin{equation}
\left( f\left( x_{1},x_{2}\right) \right) ^{t_{1}}=f\left(
x_{1},-x_{2}\right) \quad or\quad \left( f\left( x_{1},x_{2}\right) \right)
^{t_{2}}=f\left( x_{2},x_{1}\right) .  \label{anti}
\end{equation}
These two definitions are algebraically equivalent up to a redefinition of
the coordinates, $x_{\pm }=\left( x_{1}\pm x_{2}\right) /\sqrt{2}.$ Here we
have assumed a 2-dimensional noncommutative space, $\left[ x_{1},x_{2}\right]
_{\star }=i\theta ,$ for simplicity of the presentation, but the
generalization to higher dimensions is obvious, for example by promoting the
pairs of noncommuting coordinates to vectors $\vec{x}_{1},\vec{x}_{2}.$ The
functions $f$ can depend on additional commuting coordinates. These are
assumed to be present, but they will be suppressed for the sake of a simpler
presentation. These transpositions satisfy the following property under star
products
\begin{eqnarray}
\left( f\left( x\right) \star g\left( x\right) \right) ^{t_{1}} &=&\left(
g\left( x\right) \right) ^{t_{1}}\star \left( f\left( x\right) \right)
^{t_{1}}=\left( f\star g\right) \left( x_{1},-x_{2}\right) , \\
\left( f\left( x\right) \star g\left( x\right) \right) ^{t_{2}} &=&\left(
g\left( x\right) \right) ^{t_{2}}\star \left( f\left( x\right) \right)
^{t_{2}}=\left( f\star g\right) \left( x_{2},x_{1}\right) .
\end{eqnarray}
That is, the order of the functions in a product is reversed before applying
the transposition on the individual functions. As noted before, the
noncommutative functions are generically complex, but the complex number $i$
does not transform under the transpositions $t_{1}$ or $t_{2}$. This again
is similar to the rules obeyed by matrices under matrix transposition.

In fact, one can make the analogy to matrix transposition rather explicit by
recalling the matrix representation for the functions on the noncommutative
torus (using the clock-shift matrices $e^{ix_{1}}$=clock$,e^{ix_{2}}$=shift,
see e.g. \cite{nctorus}) or the noncommutative plane (e.g. see \cite{Harvey}%
). In this point of view the map $t_{1}$ is nothing but the usual
transposition in the matrix representation: the change of basis in the
matrix representation is equivalent to the coordinate reflection on the
noncommutative space. The complex number $i,$ or the sign of $\theta $ do
not change sign under matrix transposition.

To get familiar with the $t_{1}$ operation we record a few examples by
making use of the computations listed in Eq.(\ref{two}-\ref{three})
\begin{eqnarray}
\left( x_{1}\star x_{2}\right) ^{t_{1}} &=&\left( x_{1}x_{2}+\frac{i}{2}%
\theta \right) ^{t_{1}}=-x_{1}x_{2}+\frac{i}{2}\theta \\
&=&\left( x_{2}\right) ^{t_{1}}\star \left( x_{1}\right)
^{t_{1}}=-x_{2}\star x_{1}=-x_{1}x_{2}+\frac{i}{2}\theta \\
\left( x_{2}\star x_{1}\right) ^{t_{1}} &=&-\left( x_{1}\right)
^{t_{1}}\star \left( x_{2}\right) ^{t_{2}}=-x_{1}x_{2}-\frac{i}{2}\theta \\
\left( \left[ x^{\mu },x^{\nu }\right] _{\star }\right) ^{t_{1}} &=&i\theta
^{\mu \nu } \\
\left( x_{1}\star f\left( x_{1},x_{2}\right) \right) ^{t_{1}} &=&\left(
\left( x^{1}+\frac{i\theta }{2}\frac{\partial }{\partial x^{2}}\right)
f\left( x_{1},x_{2}\right) \right) ^{t_{1}}=\left( x^{1}-\frac{i\theta }{2}%
\frac{\partial }{\partial x^{2}}\right) f\left( x_{1},-x_{2}\right) \\
&=&f\left( x_{1},-x_{2}\right) \star x_{1}=\left( f\left( x_{1},x_{2}\right)
\right) ^{t_{1}}\star \left( x_{1}\right) ^{t_{1}}\,.
\end{eqnarray}
Similar exercises using $t_{2}$ are left to the reader.

We are now ready to define the set of matrices that form o$_{\star }\left(
N\right) .$ We combine the ordinary transposition of matrices, $\left(
a^{t}\right) _{ij}=a_{ji},$ with the transpositions $t_{1}$ or $t_{2}$
applied on the noncommutative functions, and denote the combined operation
with the letter $T.$ Then o$_{\star }\left( N\right) $ is given by the
antisymmetric matrices under the combined operation. Thus, for the cases $%
t_{1}$ or $t_{2}$ we demand that a matrix $a\in $ o$_{\star }\left( N\right)
$ satisfies
\begin{eqnarray}
t_{1} &:&\quad \left( \left( a\left( x_{1},x_{2}\right) \right) ^{T}\right)
_{ij}\equiv \left( a^{t}\right) _{ij}\left( x_{1},-x_{2}\right)
=a_{ji}\left( x_{1},-x_{2}\right) =-a_{ij}\left( x_{1},x_{2}\right) ,
\label{T1} \\
t_{2} &:&\quad \left( \left( a\left( x_{1},x_{2}\right) \right) ^{T}\right)
_{ij}\equiv \left( a^{t}\right) _{ij}\left( x_{2},x_{1}\right) =a_{ji}\left(
x_{2},x_{1}\right) =-a_{ij}\left( x_{1},x_{2}\right) .  \label{T2}
\end{eqnarray}
More explicitly, for the case of $t_{1}$ the parameters of o$_{\star }\left(
N\right) $ must have the form
\begin{equation}
o_{\star }(N)_{t_{1}}:a_{ij}\left( x_{1},x_{2}\right) =\left(
\begin{array}{cccc}
a_{11}\left( x_{1},x_{2}\right) & a_{12}\left( x_{1},x_{2}\right) & \cdots &
a_{1N}\left( x_{1},x_{2}\right) \\
-a_{12}\left( x_{1},-x_{2}\right) & a_{22}\left( x_{1},x_{2}\right) & \cdots
& a_{2N}\left( x_{1},x_{2}\right) \\
\vdots & \vdots & \ddots & \vdots \\
-a_{1N}\left( x_{1},-x_{2}\right) & -a_{2N}\left( x_{1},-x_{2}\right) &
\cdots & a_{NN}\left( x_{1},x_{2}\right)
\end{array}
\right)  \label{A}
\end{equation}
That is, the diagonal elements are odd functions under reflections of $%
x_{2}, $ i.e. $a_{ii}\left( x_{1},x_{2}\right) =-a_{ii}\left(
x_{1},-x_{2}\right) ,$ while the lower off-diagonal elements are related to
the upper off-diagonal elements by reflections of $x_{2}.$ Note that the
upper off-diagonal elements do not have any symmetry properties under the
reflections. Under $t_{2}$ one can define a similar o$_{\star }(N)_{t_{2}}$
whose structure differs from the one above only by replacing the reflections
of $x_{2}$ by the interchange $x_{1}\longleftrightarrow x_{2}.$

Such matrices close under the matrix-star commutation relations. Thus, for $%
a $,$b\in $ o$_{\star }\left( N\right) $, the matrix-star commutator results
in another matrix $c\in $ o$_{\star }\left( N\right) $
\begin{equation}
\left( \left[ a,b\right] _{\star }\right) _{ij}\equiv \left( a\star b-b\star
a\right) _{\,\,\,ij}=\left( c\right) _{\,\,\,ij}  \label{A1A2}
\end{equation}
In particular note that o$_{\star }(1)$ exists non-trivially. It consists of
functions that are antisymmetric under reflections of $x_{2}$ or under the
interchange $x_{1}\longleftrightarrow x_{2}$
\begin{eqnarray}
o_{\star }(1)_{t_{1}} &:&\quad a\left( x_{1},x_{2}\right) =-a\left(
x_{1},-x_{2}\right) , \\
o_{\star }(1)_{t_{2}} &:&\quad a\left( x_{1},x_{2}\right) =-a\left(
x_{2},x_{1}\right) .
\end{eqnarray}
Such \textit{odd} functions close under star commutators.

All entries in o$_{\star }(N)_{t_{1,2}}$ are generically complex functions,
however it is possible to restrict the dependence on the complex number $i$
as follows. In addition to the transposition operation $T$ of Eqs.(\ref{T1},%
\ref{T2}), we also impose the hermiticity condition as in the $u_{\star
}\left( N\right) $ case, i.e. $a^{\dagger }=-a.$ These two conditions are
compatible with each other. This shows that our o$_{\star }(N)_{t_{1,2}}$
form a subalgebra of the u$_{\star }\left( N\right) .$ The matrices $a$ that
satisfy both conditions are of the form (\ref{A}) such that the upper off
diagonal elements are of the form $a_{i<j}=a_{i<j}^{+}+is_{i<j}^{-}$ where $%
a_{i<j}^{+}$ is symmetric and $s_{i<j}^{-}$ is antisymmetric under
reflections, $a_{i<j}^{+}\left( x_{1},-x_{2}\right) =a_{i<j}^{+}\left(
x_{1},x_{2}\right) ,$ $s_{i<j}^{-}\left( x_{1},-x_{2}\right)
=-s_{i<j}^{-}\left( x_{1},x_{2}\right) ,$ while the diagonal elements $%
a_{ii}=ia_{ii}^{-}$ are purely imaginary as well as antisymmetric under
reflections. Examining only the $a_{ij}^{+}$ part of the matrix $a,$ we see
that it is real and has the same form of so$\left( N\right) $ matrices;
however these by themselves do not close under the combined star and matrix
products. Closure requires also the imaginary parameters $is_{i<j}^{-}$ and $%
is_{ii}^{-}.$ Hence, denoting an antihermitian $a$ with the symbol $a_{h},$
we may write
\begin{equation}
a_{h}=a^{+}+is^{-},  \label{ah}
\end{equation}
where $a^{+}$ is even under reflections and is an antisymmetric matrix,
while $s^{-}$ is odd under reflections and is a symmetric matrix. Both $%
a^{+} $ and $s^{-}$ are real. Note that the number of degrees of freedom in
these parameters is fewer than those in u$_{\star }\left( N\right) $ because
for u$_{\star }\left( N\right) $ the parameters are not restricted by the
reflection conditions.

Note also that the usual so$\left( N\right) $ is a subalgebra of o$_{\star
}(N)_{t_{1}}$ when all entries are functions of only $x_{1}$, i.e. $%
a_{ij}\left( x_{1}\right) ,$ since then all $s_{ij}^{-}$ vanish, and the
remaining matrix is $a^{+}$ real. For such matrices the star-matrix
commutator collapses to ordinary matrix commutator, and they obviously form
the so$\left( N\right) $ Lie algebra. Similarly so$\left( N\right) $ is a
subalgebra of o$_{\star }(N)_{t_{2}}$ when all entries are functions of only
$x_{1}+x_{2}:a_{ij}\left( x_{1}+x_{2}\right) $. This implies in particular
that the usual global $so(N)$ symmetry with the $x-$independent parameters
is the subalgebra of o$_{\star }(N)_{t_{i,2}}$.

In the following we will also need matrices $s_{ij}\left( x_{1},x_{2}\right)
$ that are symmetric under the $T$ operation, that is
\begin{eqnarray}
t_{1} &:&\quad \left( \left( s\left( x_{1},x_{2}\right) \right) ^{T}\right)
_{ij}\equiv \left( s^{t}\right) _{ij}\left( x_{1},-x_{2}\right)
=s_{ji}\left( x_{1},-x_{2}\right) =s_{ij}\left( x_{1},x_{2}\right) , \\
t_{2} &:&\quad \left( \left( s\left( x_{1},x_{2}\right) \right) ^{T}\right)
_{ij}\equiv \left( s^{t}\right) _{ij}\left( x_{2},x_{1}\right) =s_{ji}\left(
x_{2},x_{1}\right) =s_{ij}\left( x_{1},x_{2}\right) .
\end{eqnarray}
Explicitly, such matrices have the form (using $t_{1}$)
\begin{equation}
t_{1}:s_{ij}\left( x_{1},x_{2}\right) =\left(
\begin{array}{cccc}
s_{11}\left( x_{1},x_{2}\right) & s_{12}\left( x_{1},x_{2}\right) & \cdots &
s_{1N}\left( x_{1},x_{2}\right) \\
s_{12}\left( x_{1},-x_{2}\right) & s_{22}\left( x_{1},x_{2}\right) & \cdots
& s_{2N}\left( x_{1},x_{2}\right) \\
\vdots & \vdots & \ddots & \vdots \\
s_{1N}\left( x_{1},-x_{2}\right) & s_{2N}\left( x_{1},-x_{2}\right) & \cdots
& s_{NN}\left( x_{1},x_{2}\right)
\end{array}
\right)  \label{ST}
\end{equation}
Generally the entries are complex. If we impose the antihermiticity
conditions, then the diagonal is purely imaginary and even $%
s_{ii}=is^{+}\left( x_{1},x_{2}\right) ,$ while the upper off-diagonal has
the form $s_{i<j}=a_{i<j}^{-}\left( x_{1},x_{2}\right) +is_{i<j}^{+}\left(
x_{1},x_{2}\right) ,$ where $a_{i<j}^{-}\left( x_{1},x_{2}\right) $ are odd
and $s_{i<j}^{+}\left( x_{1},x_{2}\right) $ are even under reflections. That
is, the matrix elements of $s$ have opposite reflection properties to those
of the $a$ matrix discussed above. Hence, denoting an antihermitian $s$ with
the symbol $s_{h},$ we may write
\begin{equation}
s_{h}=a^{-}+is^{+}  \label{sh}
\end{equation}
where $a^{-}$ is odd under reflections and is an antisymmetric matrix, while
$s^{+}$ is even under reflections and is a symmetric matrix. Both $a^{-}$
and $s^{+}$ are real. Similar structures arise if we consider $t_{2}$
instead of $t_{1}.$ It is interesting to note schematically the matrix-star
commutation properties of these types of matrices
\begin{equation}
\left[ a,a^{\prime }\right] _{\star }\sim a^{\prime \prime },\quad \left[ a,s%
\right] _{\star }\sim s^{\prime },\quad \left[ s,s^{\prime }\right] _{\star
}\sim a.  \label{as}
\end{equation}
This closure property applies to the general complex $a,s$ as well as to the
hermitian subsets $a_{h},s_{h}.$

We now consider matrix gauge fields $A_{\mu }\left( x^{0},x_{1\alpha
},x_{2\alpha },y^{I}\right) \ \ ,\alpha =1,2,...,n,$ which depend on $\left(
x^{0},x_{1\alpha },x_{2\alpha },y^{I}\right) ,$ where $x^{0}$ is the time
coordinate, $y^{I}$ denotes commuting coordinates and $\vec{x}_{1},\vec{x}%
_{2}$ are pairs of non-commuting coordinates on which we apply the $t_{1,2}$
operations. $t_{1}$ could be applied to some components of the vectors,
while $t_{2}$ could be applied on the remaining components. For definiteness
consider only $t_{1}.$ The spacetime index takes the values $\mu =0,1\alpha
,2\alpha ,I.$ The gauge parameters are matrices $a_{h}\left( x^{0},\vec{x}%
_{1},\vec{x}_{2},y\right) $ of the form (\ref{ah}). Note that, in principle,
one can also apply reflections like $t_{1,2}$ to the commuting coordinates $%
y^{I}$. In particular, theories of this type will result in the limit $%
\theta \rightarrow 0$ for some of the non-commutative coordinates.

The 1-forms $A_{\mu }$ are matrices which transform under o$_{\star }(N)$
according to the rule
\begin{equation}
\delta A_{\mu }=\partial _{\mu }a_{h}+A_{\mu }\star a_{h}-a_{h}\star A_{\mu
}.  \label{gaugeon}
\end{equation}
Let us write these transformations explicitly for the various components $%
\mu =\left( 0,1\alpha ,I\right) \equiv m,$ and $\mu =2\alpha $%
\begin{eqnarray}
\delta A_{m} &=&\partial _{m}a_{h}+A_{m}\star a_{h}-a_{h}\star A_{m}, \\
\delta A_{2\alpha } &=&\partial _{2\alpha }a_{h}+A_{2\alpha }\star
a_{h}-a_{h}\star A_{2\alpha }.
\end{eqnarray}
The gauge fields, $A_{0},$ $A_{1\alpha },A_{I}$ have the same matrix form as
$a_{h}$ as in (\ref{ah}), as usual. However, taking into account the
reflection $\vec{x}_{2}\rightarrow -\vec{x}_{2}$ we must conclude that the
gauge field $A_{2\alpha }$ cannot be of that form. This is seen by examining
the term $\partial _{2\alpha }a$ which is not of the form $a_{h}$ after
applying the derivative, but rather it has the form $s_{h}$ as in (\ref{sh}%
). For consistency, we must also demand that $A_{2\alpha }$ is a matrix of
the form $s_{h}$ as in (\ref{sh}). Then the remainder of the transformation $%
\delta A_{2\alpha }$ is consistent according to Eq.(\ref{as}). These
observations are compatible with the coupling of the gauge field to a
current. For example, consider a particle coupled to the field $A_{\mu },$
for which the Lagrangian contains the term $\dot{x}^{\mu }A_{\mu }\left(
x\right) .$ The symmetry of the Lagrangian under the reflections $\vec{x}%
_{2}\rightarrow -\vec{x}_{2}$ requires opposite reflection properties from $%
A_{2\alpha }$ versus $A_{0},$ $A_{1\alpha },\,A_{I}$. When we consider a
string coupled to a stack of branes in the next section the same
observations will be valid.

Next, consider the gauge field strength $G_{\mu \nu }\left( x^{0},\vec{x}%
_{1},\vec{x}_{2},y\right) =\partial _{\mu }A_{\nu }-\partial _{\nu }A_{\mu
}+[A_{\mu },A_{\nu }]_{\star }$ and examine its different components
\begin{eqnarray}
G_{mn} &=&\partial _{m}A_{n}-\partial _{n}A_{m}+[A_{m},A_{n}]_{\star } \\
G_{m,2\alpha } &=&\partial _{m}A_{2\alpha }-\partial _{2\alpha
}A_{m}+[A_{m},A_{2\alpha }]_{\star } \\
G_{2\beta ,2\alpha } &=&\partial _{2\beta }A_{2\alpha }-\partial _{2\alpha
}A_{2\beta }+[A_{2\beta },A_{2\alpha }]_{\star }
\end{eqnarray}
We see from the derivative terms and from Eq.(\ref{as}) that $G_{mn}$ and $%
G_{2\beta ,2\alpha }$ have the form of $a_{h},$ while $G_{m,2\alpha }$ has
the form of $s_{h}.$ However, all components transform under the same rule
under o$_{\star }(N)$ gauge transformations (\ref{gaugeon})
\begin{equation}
\delta G_{\mu \nu }=G_{\mu \nu }\star a_{h}-a_{h}\star G_{\mu \nu }.
\end{equation}
Therefore, an invariant action for o$_{\star }(N)$ gauge theory is
\begin{equation}
S=\int dx^{0}d^{p-2n}y\int d^{2n}x\,Tr\left( G_{\mu \nu }G^{\mu \nu }\right)
,  \label{onaction}
\end{equation}
where $Tr$ is the matrix trace.

In particular, there is an o$_{\star }(1)$ gauge theory. The o$_{\star }(1)$
gauge field has fewer degrees of freedom than the u$_{\star }(1)$ gauge
field because $A_{\mu }$ can be taken as purely imaginary functions that are
odd/even under reflections (indicated by $\pm $ superscripts) $%
iA_{m}^{-}\left( x^{0},x_{1},x_{2},y\right) $ and $iA_{2\alpha }^{+}\left(
x^{0},x_{1},x_{2},y\right) .$ By contrast the u$_{\star }(1)$ gauge fields $%
A_{\mu }$ are purely imaginary functions that have no definite reflection
symmetry properties.

Let us stress that the action (\ref{onaction}) is invariant because both the
integration measure and the combination of the metric tensors $g^{\mu \nu }$
used to contract the indices of the noncommutative field strength $G_{\mu
\nu }G^{\mu \nu }$ are invariant under the reflections $t_{1,2}$ when $%
\theta ^{\mu \nu }$ is block diagonal. More generally, one may examine the
invariance of the action when $\theta ^{\mu \nu }$ is not block diagonal.
Under SO$\left( 2n\right) $ transformations in the noncommutative
directions, the form of the action does not change provided the metric $%
g^{\mu \nu }$ in $2n$ dimensions is rotationally invariant. But through the
SO$\left( 2n\right) $ rotations of the coordinates, $x^{\mu }\rightarrow
\Lambda ^{\mu \nu }x_{\nu },$ the constant $\theta ^{\mu \nu }$ gets
transformed into a new value $\left( \Lambda \theta \Lambda ^{T}\right)
^{\mu \nu }.$ It is always possible to choose $\Lambda $ such that $\Lambda
\theta \Lambda ^{T}$ is block diagonal. Thus, beginning with an action with
general $\theta ^{\mu \nu },$ and a rotationally invariant metric $g^{\mu
\nu }$, one can always transform the action to the coordinate basis in which
$\theta ^{\mu \nu }$ has a block diagonal form. This is the basis in which
we had defined the $t_{1,2}$ previously. Now we see that we can define more
general $t_{1,2}$ for general $\theta ^{\mu \nu }$ by conjugating the
previous ones with the SO$\left( 2n\right) $ rotation $\Lambda .$ This
argument shows that for general $\theta ^{\mu \nu },$ there exist some $%
t_{1,2}$ that play the same role as before. Therefore, if the metric has
rotation symmetry, then the action has the other desired symmetries for
general $\theta ^{\mu \nu }$, not only for block diagonal forms.

Finally, we point out properties of the model in the $\theta \rightarrow 0$
limit. In this ``classical limit'' of the o$_{\star }(N)$ (similarly usp$%
_{\star }(2N)$) gauge theory, the resulting commutative field theory has a
richer structure compared to the usual pure Yang-Mills theory. The outcome
depends on which quantities are held fixed as the limit is taken. First, one
may consider a straightforward $\theta \rightarrow 0$ limit in which the
gauge potentials $\left( A_{m},A_{2\alpha }\right) $ continue to have the
forms $\left( a_{h},s_{h}\right) $ of Eqs.(\ref{ah},\ref{sh}) respectively.
Then different polarizations behave differently under the transposition $T$
even in the commutative limit. However, since the imaginary parts of $\left(
a_{h},s_{h}\right) $ would no longer be needed for the closure of the
algebra, it is also possible to consider a second $\theta \rightarrow 0$
limit in which the normalization of the imaginary part is rescaled by a
power of $\theta $ before taking the limit. Then the resulting theory has
only gauge potentials $\left( A_{m}^{+},A_{2\alpha }^{-}\right) $ that are
real antisymmetric matrices (in the adjoint representation of standard so$%
\left( N\right) $), but still have respectively definite $\left(
even,odd\right) $ properties under reflections $t_{1}$, as indicated by the $%
\pm $ labels. This is still different than the usual Yang-Mills theory. Next
we notice that under dimensional compactification, in which all dependence
on $x_{2\alpha }$ is eliminated, the theory reduces to a familiar Yang-Mills
type theory, in which the $N\times N$ antisymmetric matrices in the
remaining fewer dimensions $A_{m}^{+}\left( x^{0},x_{1\alpha },y^{I}\right) $
become the standard so$\left( N\right) $ Yang-Mills gauge fields in the
adjoint representation, while the $A_{2\alpha }^{-}\left( x^{0},x_{1\alpha
},y^{I}\right) $ become scalar fields that are also in the adjoint
representation (the $\pm $ labels not needed anymore). It is possible to
achieve this reduction by considering yet a third $\theta \rightarrow 0$
limit, in which the $x_{2\alpha }$ dependence of the functions is rescaled
by $\theta $ in the form $A_{m}\left( x^{0},x_{1},\theta x_{2},y\right) $
and $A_{2\alpha }\left( x^{0},x_{1},\theta x_{2},y\right) ,$ so that $%
x_{2\alpha }$ completely disappear from the functions when $\theta $
vanishes. Finally, one may modify the last case with a $\theta \rightarrow 0$
limit that keeps both the real and imaginary parts of the scalars $%
A_{2\alpha }\left( x^{0},x_{1},y\right) =A_{2\alpha }^{-}\left(
x^{0},x_{1},y\right) +iA_{2\alpha }^{+}\left( x^{0},x_{1},y\right) $ in
which $A_{2\alpha }^{+}$ are symmetric $N\times N$ matrices.

\subsection{usp$_{\star }\left( 2N\right) $}

To define usp$_{\star }\left( 2N\right) $ we use the $t_{1,2}$
transpositions defined in the previous subsection and combine them with
matrix transposition. Then we define matrices that satisfy the condition $%
S^{T}=-CSC^{-1}$ , and $V^{T}=CVC^{-1}$, where the operation $T$ is defined
in Eqs.(\ref{T1},\ref{T2}) and the matrix $C$ is given in Eq.(\ref{C}). We
further require these matrices to be antihermitian. Then they form subsets
of the u$_{\star }\left( 2N\right) $ matrices. The matrices that satisfy
these condition have the form
\begin{equation}
S=\left(
\begin{array}{cc}
h & s \\
-s^{\dagger } & -h^{T}
\end{array}
\right) ,\quad V=\left(
\begin{array}{cc}
h^{\prime } & a \\
-a^{\dagger } & \left( h^{\prime }\right) ^{T}
\end{array}
\right) .  \label{S}
\end{equation}
Here $\left( h\left( x_{1},x_{2}\right) \right) _{\,\,j}^{i}$ or $\left(
h^{\prime }\left( x_{1},x_{2}\right) \right) _{\,\,j}^{i}$ are antihermitian
$N\times N$ matrices, as in Eq.(\ref{H}), and $\left( h\left(
x_{1},x_{2}\right) \right) ^{T}$ or $\left( h^{\prime }\left(
x_{1},x_{2}\right) \right) ^{T}$ are their transpose given by transposition
of the matrix combined with the $t_{1}$ or $t_{2}$ operation applied on the
functions of $\left( x_{1},x_{2}\right) $. Similarly, $s\left(
x_{1},x_{2}\right) $ and $a\left( x_{1},x_{2}\right) $ are independent $%
N\times N$ matrices that are symmetric/antisymmetric under the transposition
$T,$ so they have the forms of $s,a$ as in Eqs.(\ref{ST},\ref{A})
respectively, but they are not required to be antihermitian (i.e. do not
impose the additional conditions discussed following those equations).
Matrices of the type (\ref{S}), close under the star-matrix commutation
rules and form the usp$_{\star }\left( 2N\right) $ Lie algebra
\begin{equation}
\left( \left[ S_{1},S_{2}\right] _{\star }\right) _{IJ}\equiv \left(
S_{1}\star S_{2}-S_{2}\star S_{1}\right) _{\,\,IJ}=\left( S_{3}\right)
_{\,\,\,IJ}.  \label{S1S2}
\end{equation}
Evidently, this is a subalgebra of u$_{\star }\left( 2N\right) .$
Furthermore, the matrices $S,V$ close as follows
\begin{equation}
\left[ S,S^{\prime }\right] _{\star }\sim S^{\prime \prime },\quad \left[ S,V%
\right] _{\star }\sim V,\quad \left[ V,V^{\prime }\right] _{\star }\sim S.
\label{SV}
\end{equation}

One can now proceed to construct usp$_{\star }\left( 2N\right) $ gauge
theory by introducing gauge fields $A_{\mu }\left( x_{1},x_{2}\right) .$ As
in the previous section, there is a difference between $A_{0},A_{I},A_{1i}$
and $A_{2i}.$ The $A_{0},A_{I},A_{1i}$ are matrices of the form $S$ as in
Eq.(\ref{S}) while the $A_{2i}$ have the form of $V$ as in Eq.(\ref{S}).

The construction of the field strength, its consistent properties, and the
invariant action are then easily understood, as in the previous subsection,
as consequences of the closure among such matrices, Eq.(\ref{SV}).

\section{Geometric construction with D-branes}

u$_{\star }\left( N\right) $ gauge theories can be understood as a
consequence of open string theory in the presence of a constant background
magnetic field with the string boundaries ending on D-branes. When the
branes collapse on top of each other, and in the large background field
limit, the u$_{\star }\left( N\right) $ gauge theory captures the physics of
the strings and D-branes system \cite{{SW},{HD}}. What is the analog
geometric construction that corresponds to the o$_{\star }\left( N\right) $
and the usp$_{\star }\left( 2N\right) $ \ gauge theories we introduced in
the previous section?

\begin{figure}[tbp]
\centering
\includegraphics[height=8cm]{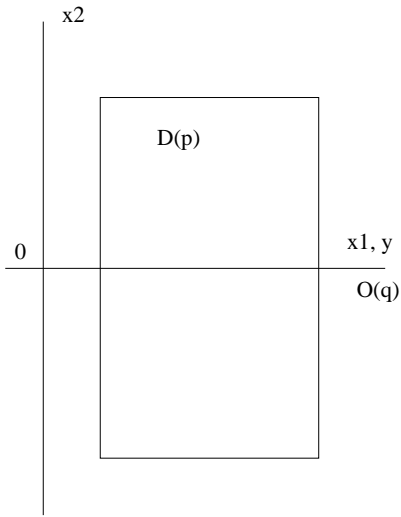} 
\caption{{D$_{p}$ brane and orientifold O$_{q}$ plane}}
\end{figure}

The answer can be found through brane configurations involving orientifolds,
chosen so that they realize the transpositions $t_{1}$ (or $t_{2})$. In
addition we also need our brane system to be supersymmetric. In order to
find a $p+1$ dimensional theory (including time $X^{0}$), let us consider $%
D_{p}$-branes which are along $1\cdots p$ spacelike directions. Let $\left(
p-2n\right) $ of them be spacelike commutative coordinates $y^{I}$ and $2n$
of them be our spacelike noncommutative coordinates $\left( x_{1\alpha
},x_{2\alpha }\right) $. More clearly, in our previous notations,

$X^{\mu },\ \ \ \ \mu =0\ $ is the time coordinate,

$\ \ \ \ \ \ \ \ \ \mu =1\alpha ,\,$with odd $\ 1\alpha =1,3,\cdots ,2n-1,\ $
are the noncommuting $x_{1\alpha }$ coordinates,

$\ \ \ \ \ \ \ \ \ \mu =2\alpha ,$ with even $2\alpha =2,4,\cdots ,2n,\ $
are the noncommuting $x_{2\alpha }$ coordinates,

$\ \ \ \ \ \ \ \ \ \mu =2n+1,\cdots ,p\ $ are the commuting $y^{I}$
coordinates,\newline
Then we choose our O$_{q}$-plane so that it has common coordinates with our D%
$_{p}$-brane in $y^{I},x_{1\alpha }$ directions, while $x_{2\alpha }$ are
transverse to the O$_{q}$-plane as in figure (1).

Also, to satisfy the supersymmetry conditions (before turning on the $B$
field), one may start from a configuration with D$_{p}$ branes and an O$%
_{p\pm 4k}$ plane. Then one may compactify one or more of the extra
dimensions on the D-brane and apply T-duality to obtain additional
supersymmetric configurations. Thus, we may start with D$_{p}+$O$_{p-4}.$
Compactifying one dimension and applying T-duality gives D$_{p-1}+$ O$_{p-3}$%
; compactifying two dimensions and applying T-duality gives D$_{p-2}$+O$%
_{p-2}$; compactifying three dimensions and applying T-duality gives D$%
_{p-3} $+O$_{p-1}$. Similarly, one could start with the systems D$_{p}+$O$%
_{p+4},$ etc. and apply the same procedure. We may shift the value of $p$ so
that we always have a D$_{p}$ brane plus an associated O$_{q}$ obtained as
above. We also note that for our scheme we need $2n\leq p\leq 9$ and $q\geq
p-n.$ This reasoning fixes the possible number of dimensions and
configurations of the D$_{p}+$O$_{q}$ of interest for our work to the
following
\begin{equation}
D_{p}+O_{p-2n+4k},\quad k=1,2.
\end{equation}
Such brane configurations will preserve $1/4$ of the 32 supersymmetries of
the type II theory. However, as we will show, the same amount of
supersymmetry remains after turning on the B-field; we will momentarily come
to this point. Supersymmetry guarantees the stability of the system from the
point of view of the complete string-brane theory, and insures that the
field theory limit makes sense as part of a finite theory.

Let us consider the D$_{p}$+O$_{q}$ configurations of interest in more
detail, in order of increasing $n$. For $n=1,$ the configuration (D$_{p}$+O$%
_{p+2})$ is described by
\begin{equation}
\begin{tabular}{llllllll}
$N\ \ D_{p}-\mathrm{branes}:$ & 0 & 1\thinspace \thinspace 2 & 3$\cdots p$ &
\quad $-$ & \quad $-$ & \quad $-$ & ,$\quad \mathrm{with}\ B_{12}$ \\
$\quad O_{p+2}-\mathrm{plane}:$ & 0 & 1$\,-$ & 3$\cdots p$ & $p+1$ & $p+2$ &
$p+3$ & .
\end{tabular}
\label{n1DpOp}
\end{equation}
where a ``$-$'' indicates that the D-brane or O-plane does not occupy the
corresponding dimensions. The reflections occur for the dimensions not
occupied by the O-plane, i.e. dimensions marked by ``$-$'' for the O-plane
(dimension $\mu =2$ in the present example). Therefore, the $B_{12}$ field
is taken with one of its indices along this dimension. This configuration
realizes a $p+1$ dimensional gauge theory on the D$_{p}$ brane worldvolume.
It has up to two noncommuting coordinates $x_{1},x_{2}$ (which become
commuting if $B_{12}=0$), a commuting time coordinate, and $p-2$ commuting $%
y^{I}$ coordinates for $2\leq p\leq 9$. We will argue below that this leads
to the gauge group o$_{\star }(N;1)$ for $N$ odd or even (or usp$_{\star
}(N;1)$ with $N$=even), where $n=1$ indicates the number of noncommuting
pairs.

In the case of $n=1$ and $p=2$ we can also consider (D$_{2}+$O$_{8}$) in a
similar way to (D$_{2}+$O$_{4}$) by adding 4 more dimensions to the O-plane
in Eq.(\ref{n1DpOp}). These additional dimensions are not occupied by the D
brane, so they do not show up in the low energy gauge theory.

For $n=2,$ the configuration (D$_{p}$+O$_{p})$ is described by
\begin{equation}
\begin{tabular}{llllllll}
$N\ \ D_{p}-\mathrm{branes}:$ & 0 & 1\thinspace \thinspace 2 & 3\thinspace
\thinspace 4 & 5$\cdots p$ & \quad $-$ & \quad $-$ & ,$\quad \mathrm{with}\
B_{12},B_{34}$ \\
$\quad O_{p}-\mathrm{plane}:$ & 0 & 1\thinspace $-$ & $3\,-$ & 5$\cdots p$ &
$p+1$ & $p+2$ & .
\end{tabular}
\end{equation}
The reflections occur for the dimensions labelled by $\mu=2\alpha =2,4$.
With $B_{12},B_{34}$ there are up to 4 non-commuting coordinates, a
commuting time coordinate, and $p-4$ commuting $y^{I}$ coordinates for $%
4\leq p\leq 9$.

For $n=3$, the configuration (D$_{p}$+O$_{p-2})$ is described by
\begin{equation}
\begin{tabular}{lllllll}
$N\ \ D_{p}-\mathrm{branes}:$ & 0 & 1\thinspace \thinspace 2 & 3\thinspace
\thinspace $4$ & 5\thinspace \thinspace 6 & 7$\cdots p$ & $\,-$\quad ,$\quad
\mathrm{with}\ B_{12},B_{34},B_{56}$ \\
$\quad O_{p-2}-\mathrm{plane}:$ & 0 & 1$\,-$ & 3\thinspace $-$ & 5\thinspace
$-$ & 7$\cdots p$ & $p+1.$%
\end{tabular}
\end{equation}
The reflections occur for the dimensions labelled by $\mu=2\alpha =2,4,6$.
With $B_{12},B_{34},B_{56}$ there are up to 6 non-commuting coordinates, a
commuting time coordinate, and $p-6$ commuting $y^{I}$ coordinates for $%
6\leq p\leq 9$.

For $n=4,$ the configuration (D$_{p}$+O$_{p-4})$ is described by
\begin{equation}
\begin{tabular}{lllllll}
$N\ \ D_{p}-\mathrm{branes}:$ & 0 & 1\thinspace \thinspace 2 & 3\thinspace
\thinspace $4$ & 5\thinspace \thinspace 6 & 7$\,\,8$ & $p$,$\quad \mathrm{%
with}\ B_{12},B_{34},B_{56},B_{78}$ \\
$\quad O_{p-4}-\mathrm{plane}:$ & 0 & 1$\,-$ & 3\thinspace $-$ & 5\thinspace
$-$ & 7$\,-$ & $p$.
\end{tabular}
\end{equation}
The reflections occur for the dimensions labelled by $\mu =2\alpha $ $=\ 2,
4, 6, 8$. With $B_{12},\ B_{34},\ B_{56},\ $ $B_{78}$ there are up to 8
noncommuting coordinates, a commuting time coordinate, and one (for $p=9$)
or zero (for $p=8$) commuting $y^{I}$ coordinates.

Although it is not indicated above, it is also possible to consider
noncommuting timelike coordinates by interchanging a spacelike coordinate
with a timelike coordinate in the above configurations and turning on
appropriate components of the $B$-field. But the field theory limit for such
configurations exists only with additional conditions (lightlike cases) as
described in \cite{gomis}.

Now consider the general constant $B$-field of rank $n$, i.e. $B_{1\alpha
\,2\beta },\ \alpha ,\beta =1,\cdots ,n$. This constant magnetic field $B$
appears in the string Lagrangian in the form
\begin{equation}
B^{1\alpha ,2\beta }\left( \partial _{\tau }X^{1\alpha }\partial _{\sigma
}X^{2\beta }-\partial _{\tau }X^{2\beta }\partial _{\sigma }X^{1\alpha
}\right) .
\end{equation}
Under the orientifold conditions it does not change sign, because there are
two sign changes applied on it, one from reflecting the field (due to $%
\sigma \leftrightarrow \left( \pi -\sigma \right) $ on the string) and one
from changing the orientation of the $X^{2\alpha }$ coordinates. This
B-field will lead to the usual noncommutativity in the $D$-brane coordinates
$x^{1\alpha },x^{2\alpha }$. This explains why, in the algebraic approach of
the previous section, the parameter $\theta $ does not change sign under
transposition.

Now that we have introduced the proper B-field, let us study the
supersymmetry of the D$_{p}+$O$_{q}$ system. It is known that in general in
the presence of the B-field the conserved supersymmetry is altered e.g. see
\cite{{Bonora},{NCUN}}. Let, $Q_{L}$ and $Q_{R}$ denote the supercharges of
corresponding type II theory. Introducing the O-plane and D-branes will
reduce the supercharges to a combination of $Q_{L}$ and $Q_{R}$, namely:
\begin{equation}
Q=\epsilon _{L}Q_{L}+\epsilon _{R}Q_{R}\ ,  \label{susy1}
\end{equation}
where $\epsilon $'s are 16 component killing spinors, and should fulfill
certain equations. As an example let us consider the brane system given by
Eq.(53). For this brane system, the $\epsilon $'s should satisfy
\begin{equation}
\Gamma ^{012\cdots p}\left( {\frac{1}{\sqrt{1+B^{2}}}}+\Gamma ^{12}{\frac{B}{%
\sqrt{1+B^{2}}}}\right) \epsilon _{L}=\epsilon _{R}\ ,  \label{sus}
\end{equation}
and
\begin{equation}
\Gamma ^{2}\Gamma ^{012\cdots p+3}\epsilon _{L}=\epsilon _{R}\ ,
\label{susy2}
\end{equation}
where $\Gamma ^{\mu }$ are ten-dimensional $16\times 16$ Dirac matrices. The
system will preserve some supersymmetry if Eqs. (\ref{sus},\ref{susy2}) have
simultaneous solutions. Solving the above for $\epsilon _{R}$ we find
\begin{equation}
A\,D(B)\epsilon _{L}=\epsilon _{L}\ ,  \label{susy3}
\end{equation}
where
\begin{equation}
A=\left( -1\right) ^{p-1}\Gamma ^{2}\Gamma ^{p+1}\Gamma ^{p+2}\Gamma ^{p+3}\
,\ \ \ \ D(B)={\frac{1}{\sqrt{1+B^{2}}}}+\Gamma ^{12}{\frac{B}{\sqrt{1+B^{2}}%
}}\ .
\end{equation}
Then we note that
\[
A^{2}=\mathbf{1}\ ,\ \ \ \ AD(B)=D(-B)A\ ,\ \ \ \ \ D(B)\ D(-B)=\mathbf{1}\
,
\]
and therefore
\begin{equation}
\left( AD(B)\right) ^{2}=\mathbf{1}\ .
\end{equation}
Also noting that Tr$\left( AD(B)\right) =0$, we conclude that the matrix $%
AD(B)$ has 8 eigenvalues equal to $+1$ and hence our brane system preserves
8 supercharges. The above argument can easily be repeated for $n=2,3,4$
cases.

The fact that our brane configuration preserves 8 supersymmetries can also
be understood if we note that upon a T-duality in the direction parallel to O%
$_{q}$-plane which also contains the B-field (e.g. $X^{1}$ direction for the
$n=1$ case of Eq.(53)), our brane system transforms to a usual D$_{p-1}$+ O$%
_{p+1}$ system without any B-field.

In the $\alpha ^{\prime }\rightarrow 0$ and $B\rightarrow \infty $ limit,
while $\alpha ^{\prime }B=\mathrm{fixed}$, we expect to find the low energy
effective theory of open strings, which should be o$_{\star }(N)$ or usp$%
_{\star }(2N)$ gauge theory in $p+1$ dimensions on the worldvolume of the D$%
_{p}$-brane. To show this and also to find the conditions on different field
polarizations, let us study the orientifold projection on the gauge fields
(massless open string states) in more detail.

As discussed in \cite{PS,GP} the orientifold projection, besides the
worldsheet parity also involves an operation on the internal degrees of
freedom (Chan-Paton factors in our case). Let, $|\psi ,\ ij\rangle $ denote
the state of an open string before the projection, where $\psi $ is the
string oscillatory (space-time) state and $i,j$ are the Chan-Paton indices.
The orientifold operation is
\begin{equation}
|\psi ,\ ij\rangle \rightarrow \gamma ^{ik}|\Omega \gamma \psi ,\ lk\rangle
(\gamma ^{lj})^{-1}
\end{equation}
where $\Omega $ is the world-sheet parity and $\gamma $ is the
representation of O-projection on the group indices. For the polarizations
of the gauge field parallel to O-plane $\Omega \gamma \psi $ is $-\psi $ and
for those which are transverse to O-plane this is $+\psi $. The above
conditions can also be written in terms of the gauge fields:
\begin{eqnarray}
A_{\mu }(y^{I},x^{1\alpha };x^{2\alpha }) &\rightarrow &\ -\gamma \ A_{\mu
}^{t}(y^{I},x^{1\alpha };-x^{2\alpha })\gamma ^{-1},\ \ \ \mathrm{for}\ \mu
\ \mathrm{parallel\ to}\ y^{I},\ x^{1\alpha }\ \mathrm{directions}  \nonumber
\\
A_{\mu }(y^{I},x^{1\alpha };x^{2\alpha }) &\rightarrow &\ +\gamma \ A_{\mu
}^{t}(y^{I},x^{1\alpha };-x^{2\alpha })\gamma ^{-1},\ \ \ \mathrm{for}\ \mu
\ \mathrm{parallel\ to}\ x^{2\alpha }\ .
\end{eqnarray}
Notice the different overall signs for different polarizations. This can be
understood from the vertex operator for gauge fields that is proportional to
$\partial X_{\mu }.$ The polarizations associated with $\partial X_{2\alpha
} $ have opposite reflection properties compared to those associated with $%
\partial X_{0},$ $\partial X_{1\alpha },$ $\partial X_{I}.$ Similar
conditions apply to the gauginos and any matter supermultiplets.

The consistency (closure) condition for the O-projection requires that
\begin{equation}
\gamma ^{-1}\ \gamma ^{t}=\pm 1\ .
\end{equation}
So, there are two possible choices, symmetric $\gamma $, which is $\gamma =%
\mathbf{1}$ for compact o$\left( N\right) ,$ and antisymmetric $\gamma $,
i.e. $\gamma =iC$ ($C$ is given in Eq.(8)), for usp$\left( N\right) $. These
two $\gamma $ choices define our O-plane: the O$_{q}^{-}$-plane corresponds
to $\gamma =\mathbf{1}$ and O$_{q}^{+}$-plane corresponds to $\gamma =iC$
\footnote{%
The usual argument that, an $O_{p}$-plane plus $D_{p+4}$-brane system, leads
to sp$(2N)$ (and not so$(N)$) \cite{GP} is basically true when the system
includes also some $D_{p}$-branes. This is not the case in our system.}.
{}From the gauge theory point of view, in fact $\gamma =\mathbf{1}$
reproduces the o$_{\star }(N)$ and $\gamma =iC$ the usp$_{\star }(N)$
algebra (for even $N$). Also we recall that $N$ is the total number of D$%
_{p} $-branes (including their reflections from the O$_{q}$-plane).

In particular we note that the o$_{\star }(1;n)$ theory is a non-trivial one
(up to $n$ noncommuting pairs), and it is obtained if we stick an O$_{q}$%
-plane to a single D$_{p}$-brane as in Fig.1.

We note that here we have just considered the case with 8 supercharges. The
classification of the supersymmetric cases and in particular the 16 SUSY
case will be studied in another paper \cite{SUSY}.

\section{Generalization of the Formalism}

It is useful to state the problem we have solved more formally in order to
provide a more general mathematical structure that could have applications
in other areas of physics. Indeed such structures have already appeared
before for classifying higher spin algebras \cite{4,3,hsr}.

First recall a few definitions. Let $B$ be some algebra with the (not
necessarily associative) product law $\diamond $. A map $\tau $ of $B$ onto
itself is called automorphism if $\tau (a\diamond b)=\tau (a)\diamond \tau
(b)$ (i.e., $\tau $ is an isomorphism of the algebra to itself.)

A useful fact is that the subset of elements $a\in B$ satisfying
\begin{equation}
\tau (a)=a  \label{taa}
\end{equation}
spans a subalgebra $B_{\tau }\subset B$. It is customary in physical
applications to use this property to obtain reductions. In particular,
applying the boson-fermion automorphism that changes a sign of the fermion
fields, one obtains reductions to the bosonic sector. Another example is
provided by the operation $\tau (a) = - a^t$ of the Lie algebra $gl(N)$ ($t$
implies transposition). The condition (\ref{taa}) then singles out the
orthogonal subalgebra $o(N)\subset gl(N)$. From the star product perspective
a collection of automorphisms of the star product algebra is provided with
the symplectic rotations of the coordinates
\begin{equation}
\tau (x^\nu ) = U^\nu{}_\mu x^\mu\,,
\end{equation}
with
\begin{equation}
U^\nu{}_\rho U^\mu{}_\sigma \theta^{\rho\sigma} = \theta^{\nu\mu}\,.
\end{equation}
In particular, one can use
\begin{equation}
\tau (x^\nu ) = - x^\nu\,.
\end{equation}
The subalgebra of the star product algebra singled out by the condition (\ref
{taa}) is spanned by the even functions
\begin{equation}
f(-x^\nu ) = f (x^\nu )\,.
\end{equation}

Let $B$ be an algebra over the field of complex numbers. If $\sigma $ is a
semilinear involutive homomorphism, i.e.
\begin{equation}
\sigma (\lambda a) = \bar{\lambda} \sigma (a)\,,\qquad \sigma (a\diamond b )
= \sigma (a) \diamond \sigma (b)\,,\qquad \sigma^2 = Id\,\qquad \forall
\lambda \in \mathbf{C} \,,\quad a,b \in B
\end{equation}
it is called conjugation. The set of elements satisfying
\begin{equation}  \label{saa}
\sigma (a) = a
\end{equation}
forms an algebra $B_\sigma$ over the field of real numbers, called real form
of $B$. For example, in a basis $\{e_i\}$ with real structure coefficients
one can define $\sigma (e_i ) = e_i $. This way one singles out, e.g., the
associative algebra of real matrices $Mat_N (\mathbf{R})$ out of $Mat_N (%
\mathbf{C})$. The same way, one can single out $gl_N (\mathbf{R})$ from $%
gl_N (\mathbf{C})$. However, for the Lie algebras there is another option
with $\sigma (a) = - (a)^\dagger$ where dagger is the Hermitian conjugation.
The resulting real Lie algebra is $u(N)$.

A linear map $\rho $ of an algebra onto itself is called antiautomorphism if
it reverses the order of product factors
\begin{equation}
\rho (a\diamond b)=\rho (b)\diamond \rho (a)\,.  \label{ant}
\end{equation}
One example is provided by the transposition of matrices. Antiautomorphisms
of the star product algebra are provided by the operations $t_{1,2}$ (\ref
{anti}).

A semilinear map $\mu$,
\begin{equation}
\mu (\lambda a) = \bar{\lambda} \mu (a)\,,\qquad \forall \lambda \in \mathbf{%
C} \,,\quad a \in B
\end{equation}
of an algebra onto itself, having the property (\ref{ant}) is called second
class antiautomorphism. If $\mu^2 =1$ we will call $\mu$ involution.
Examples of an involution are provided by the hermitian conjugation of the
matrix algebra and the operation $\dagger$ (\ref{hc}) of the star product
algebra.

Let now $A$ be some associative algebra over $\mathbf{C}$ with the product
law $f\circ g$. Let $l_{A}$ be the Lie algebra isomorphic to $A$ as a linear
space with the Lie product law defined via commutator
\begin{equation}
\lbrack f,g]=f\circ g-g\circ f\,.
\end{equation}
(For example, for $A=Mat_{N}(\mathbf{C})$, $l_{A}=gl_{N}(\mathbf{C})$).
Obviously, any automorphism, conjugation, (any class) antiautomorphism or
involution of the associative algebra $A$ acts as the operation of the same
type in the Lie algebra $l_{A}$. However, since the Lie product law is
antisymmetric, automorphisms and antiautomorphisms of Lie algebras differ
only by sign. Namely, for any antiautomorphism $\rho $ of a Lie algebra $l$,
\begin{equation}
\tau _{\rho }=-\rho
\end{equation}
is its automorphism. Analogously, an involution $\mu $ of a Lie algebra $l$
induces its conjugation
\begin{equation}
\sigma _{\mu }=-\mu \,.
\end{equation}
As a result, one can define reductions of a Lie algebra $l_{A}$ with the
help of (\ref{taa}) based both on automorphisms and antiautomorphisms of the
associative algebra $A$. Analogously, one can define real forms of $l_{A}$
using both conjugations and involutions of $A$.

If $A$ is the tensor product of two associative algebras, $A= A_1 \otimes
A_2 $, any two operations of the same type (i.e., first or second class
(anti)automorphisms) taken in combination define an operation of the same
type of $A$. We denote such combinations $\tau_1 \otimes \tau_2$, $\rho_1
\otimes \rho_2$, $\sigma_1 \otimes \sigma_2$ and $\mu_1 \otimes \mu_2$. Let
us emphasize that it is in general impossible to define a sensible operation
on $A_1 \otimes A_2$ as a combination of the operations of different types
on $A_1$ and $A_2$.

More examples are now in order. First, let $A$ be the algebra of $N\times N$
matrices over the field of complex numbers, i.e., $A=Mat_{N}(\mathbf{C})$
with elements $a^{i}{}_{j}$ ($i,j=1\div n$) and product law
\begin{equation}
(a\circ b)^{i}{}_{j}=a^{i}{}_{k}b^{k}{}_{j}\,.
\end{equation}
Let $\eta ^{ij}$ be a nondegenerate bilinear form with the inverse $\eta
_{ij}$, i.e.
\begin{equation}
\eta ^{ik}\eta _{kj}=\delta _{j}^{i}\,.
\end{equation}
It is elementary to see that the mapping
\begin{equation}
\rho _{\eta }(a)^{i}{}_{j}=\eta ^{ik}a^{l}{}_{k}\eta _{lj}  \label{antmat}
\end{equation}
is an antiautomorphism of $Mat_{N}(\mathbf{C})$. If the bilinear form $\eta
^{ij}$ is either symmetric
\begin{equation}
\eta _{S}^{ij}=\eta _{S}^{ji}  \label{sim}
\end{equation}
or antisymmetric
\begin{equation}
\eta _{A}^{ij}=-\eta _{A}^{ji}  \label{asym}
\end{equation}
the antiautomorphism $\rho _{\eta }$ is involutive, i.e. $\rho _{\eta
}^{2}=Id$. For $A=Mat_{N}(\mathbf{C})$, $l_{A}=gl_{N}(\mathbf{C})$. The
subalgebras of $gl_{N}$ singled out by the conditions (\ref{taa}) with $\tau
_{S}=-\rho _{S}$ and $\tau _{A}=-\rho _{A}$ are o$(N|\mathbf{C})$ and sp$(N|%
\mathbf{C})$, respectively, because the conditions (\ref{taa}) just imply
that the form $\eta ^{ij}$ is invariant. Analogously, one can define
involutions via nondegenerate hermitian forms. If $\dagger $ is such an
involution of $Mat_{N}(\mathbf{C})$ defined via a positive-definite
Hermitian form, then the real form of $gl_{N}(\mathbf{C})$ defined via (\ref
{saa}) with $\sigma =-\dagger $ is spanned by antihermitian matrices, thus
being u$(N)$.

Let now $A$ be the star product algebra. From the defining relations (\ref
{twop}) it follows that the definition of an involution
\begin{equation}
(x^\nu )^\dagger = x^\nu
\end{equation}
is consistent and therefore extends to the whole star product algebra. In
the particular basis associated with the Weyl (i.e. totally symmetric)
ordering, in which the star product has the Moyal form (\ref{star}), the
reordering of the operators has no effect and, therefore, the formula (\ref
{hc}) is true. Analogously, any linear map
\begin{equation}
\rho ( x^\nu ) = U(x)^\nu = U^\nu{}_\mu x^\mu
\end{equation}
such that
\begin{equation}  \label{U}
U^\nu{}_\eta U^\mu{}_\kappa \theta^{\eta\kappa} = - \theta^{\nu\mu}
\end{equation}
induces an antiautomorphism $\rho_U$ of the star product algebra. Again,
because of using the Weyl ordering, its action on a general element is
simply
\begin{equation}
\rho_U (f (x)) = f (U(x))\,.
\end{equation}
The action of the antiautomorphism on the matrix part was defined as in the
example (\ref{antmat}).

The examples of o$_{\star }$ and usp$_{\star }$ algebras given in this paper
are obtained from the application of this general scheme to the Lie algebra $%
l_{A}$ with $A=$\textit{Mat}$_{N}\otimes \mathcal{A}\quad $. For the
particular case of only two noncommutative coordinates, the map (\ref{U})
was taken in the one of the two forms
\begin{equation}
U_{1}(x^{1})=x^{1}\,,\qquad U_{1}(x^{2})=-x^{2}\,,
\end{equation}
or
\begin{equation}
U_{2}(x^{1})=x^{2}\,,\qquad U_{2}(x^{2})=x^{1}\,.
\end{equation}

The fact that the reduction was defined with the help of an automorphism of
the algebra implies that it is consistent with the definition of the
noncommutative Yang-Mills curvatures. Let us consider for definiteness the
case when all coordinates are non-commutative, i.e. $\theta ^{\mu \nu }$ is
nondegenerate, thus having inverse $\theta _{\mu \nu }$. Let us now
introduce the 1-form gauge potential
\begin{equation}
\mathcal{A}=dx^{\mu }\mathcal{A}_{\mu }=dx^{\mu }(-i\theta _{\mu \nu }x^{\nu
}I+A_{\mu }\left( x\right) )\,.
\end{equation}
where the star commutator of $-i\theta _{\mu \nu }x^{\nu }$ with any
function $f\left( x\right) $ is the derivative $\partial _{\mu }f$. The term
$dx^{\mu }\left( -i\theta _{\mu \nu }x^{\nu }I\right) $ can be treated as
the vacuum value of the potential.We obtain
\begin{equation}
\mathcal{A}\wedge \ast \mathcal{A}=dx^{\mu }\wedge dx^{\nu }(-i\theta _{\mu
\nu }I+G_{\mu \nu })\,,
\end{equation}
where $G_{\mu \nu }$ is the field strength (\ref{curv}) with the matrix
indices implicit.

If $\tau $ is some automorphism of the Lie algebra built through commutators
in \textit{Mat}$_{N}\otimes \mathcal{A}\quad $, this implies that
\begin{equation}
\tau (\mathcal{A}\wedge \ast \mathcal{A})=\tau (\mathcal{A})\wedge \ast \tau
(\mathcal{A})\,.
\end{equation}
To make it possible to truncate the system by imposing the condition
\begin{equation}
\tau (\mathcal{A})=\mathcal{A},
\end{equation}
it is necessary to insure that the vacuum value of the potential is
invariant. To this end one has to extend the action of $\tau \rightarrow
\tau ^{\prime }$ to the wedge algebra by requiring
\begin{equation}
\tau ^{\prime }(dx^{\mu }\left( -i\theta _{\mu \nu }x^{\nu }I\right)
)=dx^{\mu }\left( -i\theta _{\mu \nu }x^{\nu }I\right) ,
\end{equation}
which is possible for any $\tau $ that acts linearly on $\left( \theta _{\mu
\nu }x^{\nu }\right) $
\begin{equation}
\tau ^{\prime }(\theta _{\mu \nu }x^{\nu })=\tau (\theta _{\mu \nu }x^{\nu
})=V_{\mu }^{\,\,\nu }\left( \theta _{\nu \lambda }x^{\lambda }\right)
\end{equation}
by defining
\begin{equation}
\tau ^{\prime }(dx^{\mu })={}dx^{\lambda }\,\left( V^{-1}\right) _{\lambda
}^{\,\,\mu } \,.  \label{ty}
\end{equation}
Simultaneously, one has to redefine the action of $\tau $ on the potential
\begin{equation}
\tau ^{\prime }(\mathcal{A}_{\mu })=V_{\mu }{}^{\nu }\tau (\mathcal{A}_{\nu
})\,.  \label{taupr}
\end{equation}
As a result, the potentials can be consistently restricted by the condition
\begin{equation}
\tau ^{\prime }(\mathcal{A})=\mathcal{A},
\end{equation}
which is consistent with the field strength satisfying the similar condition
\begin{equation}
\tau ^{\prime }(\mathcal{A}\ast \wedge \mathcal{A})=\mathcal{A}\ast \wedge
\mathcal{A}\,
\end{equation}
as a consequence of the fact that $\tau ^{\prime }$ is an automorphism. The
additional signs in the transformation laws of the potentials and the field
strengths discussed in section 2 are just the particular realizations of the
definition (\ref{taupr}).

In the classical limit with commuting coordinates the associative algebra of
functions is commutative so that there is no difference between its
antiautomorphisms and automorphisms. This allows one to use the identical
(anti)automorphism of the algebra of functions in the standard construction
of the usual (i.e., commutative non-Abelian) Yang-Mills theory. In the
non-commutative case this is not allowed any longer. As a result, the
classical limit of our o${}_{\star }$ and usp${}_{\star }$ noncommutative
Yang-Mills theories is different from the usual Yang Mills theory, because
its matrix content is dependent on the oddness of the Yang-Mills potentials
as in (\ref{A}), as discussed at the end of section 2.2.

Let us note that in \cite{4,3} where similar technics were originally
applied to the tensor product of the star product algebra with matrix
algebras, relevant to the problem of higher spin gauge fields \cite{hsr}, it
was discussed in a more general framework of Lie superalgebras rather than
Lie algebras. For more detail on the relationships between (semi)linear
(anti)automorphisms associative algebras and real forms and reductions of
the associated Lie superalgebras built via (anti)commutators we therefore
refer the reader to the first reference in \cite{4}.

\section{Outlook}

In this work we have studied the formulation of noncommutative o$_{\star
}(N) $ and usp$_{\star }(2N)$ algebras and the corresponding gauge theories.
Our method is based on the specific realization of an antiautomorphism $\rho$
that changes the order of the functions in the star products, Eq. (\ref
{rhomap}). We showed that the map $\rho $ can be obtained through an
operation which acts on the noncommutative coordinates, the $t_{1}$ or $%
t_{2} $ operations. In this way we can relax the $\theta $ dependence of
functions which were assumed in \cite{Bonora} for the representation of this
map. The $t_{1,2}$ transformations do not change $\theta $.

Then we discussed how the combination of the $t_{1}$ (or $t_{2}$) with the
usual matrix transposition, the $T$ operation, provides the proper
``transposition'' for the general star-matrix algebras. Particularly
starting with the u$_{\star }(N)$ algebra, and restricting it to the
anti-symmetric elements under $T$ operation, leads to o$_{\star }(N)$ as a
subalgebra. Similarly one can construct usp$_{\star }(2N)$ as a subalgebra
of u$_{\star }$ $\left( 2N\right) .$ In order to formulate o$_{\star }(N)$
field theories, including scalar, fermion and vector fields, we studied the
proper representations and showed that for the gauge fields, the $T$
operation also imposes some conditions on the polarizations which are not
the same in every direction.

We noted that in the $\theta \rightarrow 0$ limit the noncommutative
Yang-Mills theory does not necessarily reduce to the standard Yang-Mills
theory if there is no reduction in the number of dimensions. This is because
of the twisted symmetry properties of the different polarizations of the
gauge fields under the transposition $T$. However, one may consider several
different types of $\theta \rightarrow 0$ limits in which various quantities
are held fixed. In particular one may consider the limit in which the $%
x_{2\alpha }$ coordinates are dimensionally reduced. Then the resulting
limiting theory is a Yang Mills theory in lower dimensions with additional
scalars in various representations.

We also discussed the brane configurations with 8 supercharges which lead to
the o$_{\star }(N)$ and usp$_{\star }(2N)$ gauge theories. We discussed how
exactly the same conditions and requirements that we considered from an
algebraic point of view also follows from string theory.

The discussion of other D$_{p}+$O$_{q}$ configurations in the presence of
the $B$ field with other amounts of supersymmetry, and the structure of the
corresponding supersymmetric o$_{\star }(N)$ or usp$_{\star }\left(
2N\right) $ gauge theories are left for a future paper \cite{SUSY}.

Let us mention some possible areas where our results could be useful. One
potential application of the o$_{\star }(N)$ group is in the context of
noncommutative gravity. Previously, because a noncommutative version of so$%
(N)$ was lacking, noncommutative gravity was attempted by gauging $u_{\star
}(3,1)$ \cite{Chams}. Among other challenges of this approach, $u_{\star
}(3,1)$ has the disadvantage that the gravity field, the metric, becomes a
complex field. However, it seems more plausible to attempt a construction of
noncommutative gravity by gauging o$_{\star }(3,1),$ which is an analytic
continuation of the o$_{\star }(4)$ we discussed here.

Another interesting open problem one could address for the 16 SUSY case is
the corresponding noncommutative open string (NCOS) theory \cite{NCOS}.
Usually the NCOS appears in noncommutative space-times, in the critical
noncommutativity limit (when the noncommutativity scale and the string scale
become identical.) In our case, we expect that the same critical limit
exists and will lead to {unoriented} NCOS \cite{prog}.

It is known that the $D=4,\ N=2$ theories can be studied through the
holomorphic Seiberg-Witten curve \cite{N=2}. On the other hand it has been
shown that the same curve can be understood more intuitively through brane
configurations involving type IIA $NS5$-branes \cite{Witten}. For the u$%
_{\star }(N)$ theories the corresponding brane configuration have been
discussed \cite{NCUN}. Since adding the $NC5$-brane and the orientifolds do
not break any further supersymmetries \cite{braneconfig} we expect that the o%
$_{\star }(N)$ theories can be studied through the ``curve'' obtained from
our brane configurations (of course after adding the needed $NS5$-branes).

As a first extension of our work one may look for the superalgebraic
generalization of the o$_{\star }(N)$ and usp$_{\star }(2N)$ algebras.

\section*{Acknowledgments}

We gratefully acknowledge very helpful discussions with S. Sugimoto who
provided us with useful information on the D brane and orientifold
geometries mentioned in this paper. In addition, I.B. and M.V. would like to
thank D. Minic, R. Corrado and E. Witten, and M.M.Sh-J would like to thank
L. Bonora, K. Narain and A. Tomasiello, for fruitful discussions. M.V. would
like to thank the CIT-USC Center for Theoretical Physics for hospitality
where this work was performed. I.B. and M.M.Sh-J would like to thank the 37$%
^{th}$ Karpacz Winter School for hospitality where discussions on this work
was initiated. I. Bars was partially supported by the US Department of
Energy under grant number DE-FG03-84ER40168, and the CIT-USC Center for
Theoretical Physics. M.M. Sh-J. was partly supported by the EC contract no.
ERBFMRX-CT 96-0090. M.V. was partially supported by INTAS, Grant No.99-1-590
and by the RFBR Grant No.99-02-16207.

\end{document}